\documentclass[journal=jctcce,manuscript=article]{achemso}
\usepackage{graphicx} 
\usepackage{amsmath} 
\usepackage{braket}
\usepackage{bm}
\usepackage{glossaries}
\usepackage[dvipsnames]{xcolor}
\usepackage{booktabs}
\usepackage{amssymb} 
\usepackage{subcaption}
\usepackage{orcidlink}
\usepackage{multirow}
\usepackage{hyperref}
\hypersetup{colorlinks,
    citecolor=blue,
    linkcolor=red,
    urlcolor=purple,
    breaklinks=true}

\usepackage{mhchem}
\newcommand{\ndf}{N_{\text{DF}}}

\newacronym{ci}{CI}{configuration interaction}
\newacronym{cc}{CC}{coupled cluster}
\newacronym{qpe}{QPE}{quantum phase estimation}
\newacronym{lcu}{LCU}{linear combination of unitaries}
\newacronym{mo}{MO}{molecular orbital}
\newacronym{ao}{AO}{atomic orbital}
\newacronym{gto}{GTO}{Gaussian-type orbital}
\newacronym{hf}{HF}{Hartree-Fock}
\newacronym{fci}{FCI}{full configuration interaction}
\newacronym{ano}{ANO}{atomic natural orbital}
\newacronym{fno}{FNO}{frozen natural orbital}
\newacronym{no}{NO}{natural orbital}
\newacronym{df}{DF}{double factorization}
\newacronym{thc}{THC}{tensor hypercontraction }
\newacronym{rdm}{RDM}{reduced density matrix}
\newacronym{sci}{SCI}{selected configuration interaction}
\newacronym{mp2}{MP2}{second-order Møller–Plesset perturbation}
\newacronym{dmrg}{DMRG}{density matrix renormalization group}

\title{Improving the runtime of quantum phase estimation for chemistry through basis set optimization}

\author{Pauline J. Ollitrault\orcidlink{0000-0003-1351-7546}}
\email{pauline.ollitrault@qcware.com}
\affiliation[QCW]{QC Ware Corporation, Palo Alto, California 94301, USA}

\author{J\'er\^ome F. Gonthier\orcidlink{0000-0002-2933-4085}}
\affiliation[QCW]{QC Ware Corporation, Palo Alto, California 94301, USA}

\author{Dario Rocca\orcidlink{0000-0003-2122-6933}}
\affiliation[QCW]{QC Ware Corporation, Palo Alto, California 94301, USA}

\author{Gian-Luca Anselmetti\orcidlink{0000-0002-8073-3567}}
\affiliation[BI]{Quantum Lab, Boehringer Ingelheim, 55218 Ingelheim am Rhein, Germany}

\author{Matthias Degroote\orcidlink{0000-0002-8850-7708}}
\affiliation[BI]{Quantum Lab, Boehringer Ingelheim, 55218 Ingelheim am Rhein, Germany}

\author{Nikolaj Moll\orcidlink{0000-0001-5645-4667}}
\affiliation[BI]{Quantum Lab, Boehringer Ingelheim, 55218 Ingelheim am Rhein, Germany}

\author{Raffaele Santagati\orcidlink{0000-0001-9645-0580}}
\affiliation[BI]{Quantum Lab, Boehringer Ingelheim, 55218 Ingelheim am Rhein, Germany}

\author{Michael Streif\orcidlink{0000-0002-7509-4748}}
\affiliation[BI]{Quantum Lab, Boehringer Ingelheim, 55218 Ingelheim am Rhein, Germany}

\begin{document}

\begin{abstract}
Quantum phase estimation (QPE) is a promising quantum algorithm for obtaining molecular ground-state energies with chemical accuracy. However, its computational cost, dominated by the Hamiltonian 1-norm $\lambda$ and the cost of the block encoding, scales at least quadratically with the number of molecular orbitals, making it challenging to incorporate dynamic correlation by enlarging the active space. In this work, we investigate two strategies to mitigate this cost through the optimization of the basis set.  First, we investigate whether adjusting the coefficients of Gaussian basis functions can minimize the 1-norm while preserving the accuracy of the ground state energy. Although this method leads to a reduction in the 1-norm up to 10\%, this reduction is system-dependent and diminishes with increasing molecular size. Second, we demonstrate that employing a large-basis-set frozen natural orbital (FNO) strategy results in a substantial reduction in QPE resources without compromising accuracy. We study a dataset of 58 small organic molecules and the dissociation curve of \ce{N2}, and demonstrate that an active space constructed from orbitals derived from larger basis sets captures correlation effects more effectively. This approach yields up to an 80\% reduction in the 1-norm $\lambda$ and also leads to a 55\% reduction in the number of orbitals. Our results highlight that improving the quality, not just the size, of the orbital basis is a viable strategy for extending QPE to include dynamical correlation, making progress toward scalable and chemically accurate quantum simulations with tractable resource requirements.
\end{abstract}

	\maketitle

\section{Introduction}
Quantum chemistry is essential for understanding and predicting the behavior of molecules, which underpins advances in drug discovery, catalysis, and environmental chemistry. However, solving the Schrödinger equation for systems with many interacting electrons with traditional computers requires immense computational power -- growing exponentially with system size. Quantum computing operates on quantum bits and can naturally simulate quantum states, offering an attractive alternative to solve the Schrödinger equation more efficiently~\cite{reiher2017elucidating, cao2019quantum, bauer2020quantum}.

The \gls{qpe} algorithm is the prototypical algorithm for computing the ground state energy of an electronic molecular Hamiltonian $\hat{H}$, on quantum computers.~\cite{abrams1997simulation, abrams1999quantum, aspuru2005simulated} Recent advances in reducing the resources required to run \gls{qpe} have been driven by the qubitization method~\cite{low2019hamiltonian, babbush2018encoding}, which reformulates the problem in terms of a walk operator $W[\hat{H}]=e^{-i \arccos{(\hat{H}/\lambda})}$. QPE then effectively returns the arccosine of the ground state energy. Qubitization requires the Hamiltonian to be expressed as a \gls{lcu}, where the coefficients are normalized with the 1-norm $\lambda$. The cost of the qubitization-based \gls{qpe} scales as $\mathcal{O}\big(\frac{\lambda}{\epsilon_{\text{QPE}}}C_{W}\big)$, where $\epsilon_{\text{QPE}}$ is the accuracy required in the resulting energy, and $C_{W}$ is the cost of implementing the walk operator, which depends on the number of terms in the \gls{lcu} decomposition.~\cite{low2024trading, berry2019qubitization}.

Significant progress has been made to find improved \gls{lcu} representations of the Hamiltonian to reduce both $\lambda$ and $C_W$. Techniques like \gls{df}~\cite{von2021quantum, motta2021low, cohn2021quantum, oumarou2024accelerating} and \gls{thc}~\cite{lee2021even,goings2022reliably} for matrix factorization, as well as symmetry shifting methods~\cite{loaiza2023reducing, loaiza2023block,rocca2024reducing,caesura2025faster} have led to substantial resource savings. More recently, spectral amplification has been proposed to reduce the linear scaling of \gls{qpe} on $\lambda$ to square root dependence~\cite{low2025fast}. For example, the estimated runtime of \gls{qpe} for a 76 active orbital model of FeMoco -- designed to capture static correlation -- was reduced from 12 days~\cite{lee2021even} to less than 9 hours~\cite{low2025fast}.

However, estimating ground state energies of systems with significant dynamic correlation, where accurate energy predictions require accounting for contributions from orbitals outside the active space, remains challenging. 
Various methods have been proposed to address this, including perturbation theory~\cite{mcclean2016theory, tammaro2023n, krompiec2022strongly}, subspace expansion~\cite{takeshita2020increasing, tammaro2023n, baek2023say}, embedding approaches~\cite{rossmannek2021quantum, ma2021quantum, li2022toward, liu2023bootstrap, he2020zeroth, he2022second, evangelista2018perspective, huang2023leveraging}, or wavefunction corrections~\cite{scheurer2024tailored}. These methods typically depend on high-order \gls{rdm}, which are prohibitively expensive to obtain within QPE due to the deep circuits involved.

The most natural way to incorporate dynamic correlation in \gls{qpe} is to expand the active space by including more orbitals. However, this strategy poses significant challenges: it demands more qubits, a greater number of terms in the \gls{lcu} decomposition of the Hamiltonian, and the 1-norm of the Hamiltonian typically grows at least quadratically with the number of orbitals. Consequently, running QPE beyond small active spaces toward the complete basis set limit becomes seemingly impractical due to the prohibitive computational cost.

In this work, we investigate whether the choice of basis set can be fine-tuned to mitigate the growth of the 1-norm while preserving the accuracy of computed energies. As a first step, we directly optimize the exponent and contraction coefficients of the Gaussian functions composing the basis set. This strategy yields only modest improvements of up to 10 \%. In contrast, our central result shows that the cost of \gls{qpe} can be substantially reduced by employing the \gls{fno}~\cite{barr1970nature} strategy, starting from a large basis set. In the \gls{fno} approach, part of the virtual orbital space is truncated, thereby lowering the cost of the subsequent quantum algorithm while still including the dynamical correlation. We show that deriving \glspl{fno} from a dense basis set yields significantly greater savings compared to constructing them from a smaller basis. This result suggests that coarse basis sets should be avoided, and cost reductions are more effectively achieved through the \gls{fno} strategy.

In Section~\ref{sec:theory}, we
present the theoretical background. We review the theory of Hamiltonian block-encoding and define the operator norms used throughout this work. We further provide an overview of atomic basis sets, which form the foundation for our basis set optimization methodology, and introduce the \gls{fno} approach. Numerical simulations and their analysis are presented in Section~\ref{sec:results}, and conclusions and perspectives are given in Section~\ref{sec:conclusion}.

\section{Theoretical background}
\label{sec:theory}

\subsection{Hamiltonian block-encoding, sparse and LCU norms}
\label{sec:hamiltonian_block-encoding}

The molecular electronic Hamiltonian, in the second quantization formalism, is expressed as
\begin{equation}
    \hat{H} = \hat{T} + \hat{V}
\end{equation}
with
\begin{equation}
    \hat{T} = \sum_{\sigma \in \{\uparrow,\downarrow\}} \sum_{p,q=1}^{N} T_{pq} \, \hat{a}_{p,\sigma}^\dagger \hat{a}_{q,\sigma}
\end{equation}
and 
\begin{equation}
    \hat{V} = \frac{1}{2} \sum_{\alpha,\beta \in \{\uparrow,\downarrow\}} \sum_{p,q,r,s=1}^{N} V_{pqrs} \, \hat{a}_{p,\alpha}^\dagger \hat{a}_{q,\alpha} \hat{a}_{r,\beta}^\dagger \hat{a}_{s,\beta} \, ,
\end{equation}
where $N$ is the number of \glspl{mo}. The one- and two-electron integrals are
\begin{equation}
    V_{pqrs} = \iint \mathrm{d}\mathbf{r}_1 \, \mathrm{d}\mathbf{r}_2 \, 
\frac{ \psi_p(\mathbf{r}_1) \psi_q(\mathbf{r}_1) \psi_r(\mathbf{r}_2) \psi_s(\mathbf{r}_2) }{ |\mathbf{r}_1 - \mathbf{r}_2| }
\end{equation}
and
\begin{equation}
    h_{pq} = \int d\mathbf{r} \, \psi_p(\mathbf{r}) 
\left( -\frac{1}{2} \nabla^2 - \sum_I \frac{Z_I}{r_I} \right) 
\psi_q(\mathbf{r}) \, ,
\end{equation}
where $\psi_p(\mathbf{r})$ are the single-particle \glspl{mo}, $\nabla^2$ is the Laplacian operator (kinetic energy), $Z_I$ is the nuclear charge of nucleus $I$, and $r_I$ is the distance between electron and nucleus $I$.
Finally, the modified one-body tensor is defined as
\begin{equation}
    T_{pq} = h_{pq} - \frac{1}{2} \sum_{r=1}^{N} V_{prrq} \, .
\end{equation}

To use the second-quantized Hamiltonian in a quantum computing context, it must be expressed as a linear combination of unitary operators. The most common and conceptually straightforward approach is to map the fermionic creation and annihilation operators, $\hat{a}_{p,\alpha}^\dagger, \hat{a}_{q,\alpha}$, to qubit operators via the Jordan–Wigner transformation. This procedure transforms the Hamiltonian into a linear combination of Pauli strings $\hat{P}_k \in \{ \hat{I}, \hat{X}, \hat{Y}, \hat{Z}\}^{\otimes N}$, i.e.,
\begin{equation}
    \hat{H}_Q = \sum_k h_k \hat{P}_k.
\end{equation}
The 1-norm associated with this decomposition is then
\begin{equation}
    \lambda_Q = \sum_k |h_k|
\end{equation}
and scales with the number of terms in $\hat{H}_Q$, i.e. $\mathcal{O}(N^4)$. There is some freedom in the fermion to qubit mapping and the exact value of 1-norm depends on the grouping of certain terms. Koridon \textit{et al.}~\cite{koridon2021orbital} demonstrated that this norm can be expressed from the one and two-electron integrals as follows
\begin{equation}
    \lambda_Q = \lambda_C + \lambda_T + \lambda_V \, ,
\end{equation}
where
\begin{align}
    &\lambda_C = \left| \sum_{p}^{N} h_{pp} + \frac{1}{2} \sum_{pr}^{N} V_{pprr} - \frac{1}{4} \sum_{pr}^{N} V_{prrp} \right|,\\
    & \lambda_T = \sum_{pq}^{N} \left| h_{pq} + \sum_{r}^{N} V_{pqrr} - \frac{1}{2} \sum_{r}^{N} V_{prrq} \right|, \\
    & \lambda_V = \frac{1}{2} \sum_{\substack{p > r,\, s > q}}^{N} \left| V_{pqrs} - V_{psrq} \right| + \frac{1}{4} \sum_{pqrs}^{N} \left| V_{pqrs} \right| \, .
\end{align}
This expression is useful as it allows us to isolate the first term $\lambda_C$, which corresponds to the absolute value of the coefficient of the identity term after the Jordan-Wigner transformation. Since this term contributes only to a global phase, it does not affect the dynamics of the quantum state and therefore does not need to be implemented on a quantum computer. Instead, its effect can be incorporated through classical post-processing. As a result, the effective 1-norm that determines the cost of the quantum algorithm becomes
\begin{equation}
    \lambda_Q^{'} = \lambda_T + \lambda_V \, .
    \label{eq:sparse_norm}
\end{equation}

To reduce the $\mathcal{O}(N^4)$ complexity, several tensor factorization techniques have been developed. One widely used approach is explicit \gls{df}, which involves performing two successive Cholesky decompositions of the two-electron integral tensor. This yields the approximate form:
\begin{equation}
    V_{pqrs} \approx \sum_{t=1}^{\ndf} \sum_{kl}^N U^t_{pk} U_{qk}^t V_{kl}^t U_{rl}^t U_{sl}^t \, ,
    \label{eq:df_2electron_tensor}
\end{equation}
where each tensor $\bm{U}^t$ tensor is orthonormal, and the core tensors $\bm{V}^t$ are positive semi-definite and of rank one for all $t$. Consequently, each core tensor can be further factorized as
\begin{equation}
    V^{t}_{kl} = W^{t}_k W^{t}_l.
\end{equation}
The value $\ndf \leq N^2$ is chosen to balance accuracy and computational efficiency. Within this representation, the 1-norm dictating the quantum algorithm cost becomes~\cite{von2021quantum, lee2021even, rocca2024reducing},
\begin{equation}
    \lambda_{\text{DF}} = \sum_k^N |f^{\varnothing}_k| + \frac{1}{4} \sum_t^{\ndf}\Big( \sum_k^N |W_k^t|\Big)^2 \, ,
    \label{eq:df_norm}
\end{equation}
where the one-electron tensor is redefined as
\begin{equation}
    f_{pq} = T_{pq} + \sum_r^N V_{pqrr}
\end{equation}
and (single) factorized as
\begin{equation}
    f_{pq} = \sum_k^N U^{\varnothing}_{pk} f^{\varnothing}_{k}U^{\varnothing}_{qk} \, .
\end{equation}

Other techniques, such as compressed and regularized \gls{df}~\cite{oumarou2024accelerating, cohn2021quantum}, \gls{thc}~\cite{lee2021even}, and their symmetry-aware variants~\cite{rocca2024reducing,caesura2025faster}, can achieve even smaller effective 1-norms. A key feature shared by all these methods is their invariance under orbital rotations, which is the property of primary importance for our purposes. Among them, explicit \gls{df} offers a more computationally efficient route to evaluating 1-norms. Therefore, we adopt this approach throughout the present work.

\subsection{Atomic Orbitals}
\label{sec:atomic_basis_function}

In most quantum chemistry methods, electronic wave functions are constructed from antisymmetrized products of \glspl{mo}, each of which is expanded in terms of a finite set of basis functions. The choice of these basis functions has a critical impact on both the accuracy and the computational cost of electronic structure calculations.

In molecular calculations, \glspl{mo} are commonly expressed as linear combinations of \glspl{ao},
\begin{equation}
\psi_i(\mathbf{r}) = \sum_\mu C_{\mu i} \phi_\mu(\mathbf{r}) \, ,
\end{equation}
where \( \phi_\mu \) denotes an \gls{ao} function centered on a nucleus and \( C_{\mu i} \) is a \gls{mo} coefficient. In the most general form, \glspl{ao} are typically represented as a fixed linear combination of basis functions $\chi_{\mu}^{\alpha}(\mathbf{r})$, each being a product of a radial function and an angular function:
\begin{eqnarray}
\phi_{\mu}(\mathbf{r}) &=& \sum_{\alpha} d_{\mu}^{\alpha}\chi_{\mu}^{\alpha}(\mathbf{r}) \\
 &=& \sum_{\alpha} d_{\mu}^{\alpha} R^{\alpha}_{nl}(r) Y^{\alpha}_{lm}(\theta, \phi) \, ,
\end{eqnarray}
where $n$, $l$, $m$ are angular quantum numbers, \( R^{\alpha}_{nl}(r) \) is the radial component and \( Y^{\alpha}_{lm}(\theta, \phi) \) is a spherical harmonic determining the angular shape. Note that here, we assume that $n$, $l$, and $m$ only depend on $\mu$ for simplicity. 

In computational quantum chemistry, a \textit{basis set} refers to the finite collection of \glspl{ao} whose coefficients can be varied independently to build \glspl{mo}. The basis set serves as a discrete representation of the infinite-dimensional Hilbert space of square-integrable one-electron functions. Different basis sets vary essentially in the radial part of each \glspl{ao}, which can be represented with different basis functions or combinations of basis functions; and in the total number of \glspl{ao} available to build \glspl{mo}. The overall flexibility and accuracy of the molecular wavefunction are constrained by the completeness and quality of the basis set. Therefore, the construction of \glspl{ao} critically influences the accuracy of any quantum chemical method, as it depends critically on this representation.

Selecting good basis functions and building good \glspl{ao} involve trade-offs. Ideally, selected basis functions should offer systematic convergence toward completeness within the space of square-integrable one-electron functions, enable rapid convergence for both atomic and molecular states, and admit an analytic form suitable for efficient integral evaluation. In practice, it is challenging to fulfill all these requirements simultaneously, and practical basis sets are constructed to strike a balance between flexibility, compactness, and numerical stability.

\subsubsection{Gaussian Basis Functions}

\Glspl{gto} are the most commonly used basis functions in molecular quantum chemistry. Although they do not correctly reproduce the cusp behavior at the nucleus or the exponential decay of atomic orbitals, they offer a significant computational advantage: integrals over \glspl{gto} can be evaluated analytically and very efficiently.

A normalized Cartesian GTO centered at a position \( \mathbf{R}_A \) has the general form
\begin{equation}
\chi(\mathbf{r}) = Y_{lm}^{GTO}(\theta, \phi) e^{-\alpha |\mathbf{r} - \mathbf{R}_A|^2} \, ,
\end{equation}
where $\alpha$ is the exponent controlling the radial extent. Despite requiring a relatively large number of \glspl{gto} to achieve convergence of the Hartree--Fock ground state, this does not pose a severe problem for integral evaluation due to the efficiency of the \gls{gto} form. However, the cost of correlated methods rapidly increases with the size of the basis set, i.e. the total number of \glspl{ao}, therefore it is quite inefficient to build each \gls{ao} out of a single \gls{gto}.

To reduce the number of \glspl{ao} while preserving accuracy, many \glspl{gto} are typically combined into a single \gls{ao} in a so-called contracted Gaussian: a fixed linear combination of \glspl{gto} with predetermined coefficients. A contracted Gaussian function is written as
\begin{equation}
\label{eq:contraction}
\phi_{\mu}(\mathbf{r}) = \sum_{i} d_{\mu}^{i} Y_{lm}^{GTO}(\theta, \phi) e^{-\alpha_{i} |\mathbf{r} - \mathbf{R}_A|^2} \, . 
\end{equation}
Depending on how the \glspl{gto} are combined, contractions may be either segmented, with each \gls{gto} used in only one contracted function (as in Pople-type basis sets), or general, where \glspl{gto} can contribute to multiple contracted orbitals (as in ANO and correlation-consistent sets).

The concept of \emph{zeta} quality refers to the number of \glspl{ao} of the same angular momentum but with different radial functions used to describe a given valence orbital. In a double-zeta (DZ) basis set, each valence orbital is described by two \glspl{ao}, providing more flexibility than a single-zeta basis. Triple-zeta (TZ) and quadruple-zeta (QZ) sets use three and four \glspl{ao} per valence orbital, respectively. Increasing the zeta level improves the basis set's ability to describe changes in the electronic environment and electron correlation effects.

\subsubsection{Atomic Natural Orbital Basis Sets}

For correlated electronic structure methods, the quality of the virtual orbital space is essential, as it determines the extent to which electron correlation can be captured. \Gls{ano} basis sets are constructed to offer a highly compact yet flexible description of both occupied and virtual spaces. Their construction begins with a Hartree--Fock calculation on an atomic system using a large primitive Gaussian basis set, yielding a set of canonical orbitals. 
A correlated atomic calculation (typically CISD) is then performed in the full space of canonical orbitals. From this correlated wavefunction, the one-electron density matrix is constructed and diagonalized in the virtual subspace. The eigenfunctions, i.e., natural orbitals, are ranked by occupation number, and those with the highest occupations are retained. These correlating orbitals, together with the original occupied canonical orbitals, form the \gls{ano} basis set.

By selecting natural orbitals based on their occupation numbers, one obtains a systematically improvable and hierarchically truncatable basis set. \glspl{ano} offer excellent accuracy per \gls{ao}. However, the resulting basis sets often involve large numbers of primitive Gaussians, particularly when high angular momentum functions are required. 

\subsubsection{Correlation-Consistent Basis Sets}

Correlation-consistent basis sets, introduced by Dunning, are designed to reproduce the efficiency of \gls{ano} sets while avoiding their computational complexity. These basis sets are constructed by optimizing the exponents of \glspl{gto} to maximize their contribution to the correlation energy. Functions are grouped by angular momentum, and each new “shell” of functions is added only when it contributes significantly to the energy. This yields a hierarchy of basis sets in which correlation effects are recovered in a controlled and systematic fashion.

A key feature of correlation-consistent basis sets is that each set includes all functions that contribute comparably to the total correlation energy, leading to a smooth convergence with increasing basis set size. The resulting series, denoted as cc-pV$n$Z, where \( n \) indicates the zeta level, includes sets such as cc-pVDZ (double-zeta), cc-pVTZ (triple-zeta), cc-pVQZ (quadruple-zeta), etc. Each level adds higher angular momentum functions (e.g., \( d \), \( f \), \( g \)) to increase the flexibility of the basis and improve the treatment of electron correlation.

Interestingly, the sequence of correlating orbitals selected by the energy-based optimization in correlation-consistent basis sets closely mirrors that obtained from \glspl{ao} based on occupation-number criteria. However, the energy-based construction requires fewer \glspl{ao} to achieve similar accuracy, making correlation-consistent basis sets particularly well suited for routine molecular calculations.

\subsection{Frozen natural orbitals}
\label{sec:fnos}

The concept of natural orbitals was introduced by Löwdin in 1955~\cite{lowdin1955quantum}, with early application to orbital space reduction by Barr and Davidson~\cite{barr1970nature} and Sosa et al.~\cite{sosa1989} laying the groundwork for space-reduction techniques in \gls{ci} and \gls{mp2} calculations, respectively. A major advancement was made by Taube and Bartlett, who formalized the \gls{fno} approach for \gls{cc}~\cite{taube2005, taube2008} showing that substantial reductions in virtual orbital space can be achieved with minimal loss in correlation energy~\cite{deprince2013, deprince2013b, gyevi2021, nagy2021}.

In quantum computing, \glspl{fno} have gained renewed relevance due to the limitations on qubit and gate counts. By compactly representing the virtual space, they reduce active spaces for quantum algorithms~\cite{verma2021, mochizuki2019}. 

Here, \glspl{fno} are employed in a slightly modified manner where the truncation technique is used to compare orbitals derived from two different basis sets. By applying different truncation thresholds, we obtain two orbital sets of equal size, resulting in equivalent resource estimates (e.g., comparable $\lambda$ and $C_W$ values). However, as we will show, these orbital sets yield different correlation energies. Notably, orbitals derived from a more refined basis set capture correlation effects more effectively, even at higher truncation thresholds.

\section{Results}
\label{sec:results}

We pursued two complementary approaches to investigate the effect of the basis set on the \gls{qpe} cost: in Section~\ref{sec:res:basis_optimization} we directly optimize the Gaussian basis functions using a cost function that accounts for the 1-norm of the Hamiltonian, and in Section~\ref{sec:res:fno_strategy} we compare the \glspl{fno} generated from different basis sets. Unless stated otherwise, all electronic structure calculations in this work are performed using PySCF~\cite{sun2018pyscf, sun2020recent}.

\subsection{Optimization of the atomic orbital basis}
\label{sec:res:basis_optimization}

To reduce the cost of QPE, we explore in this section whether changing the structure of Gaussian basis sets can reduce the 1-norm of the Hamiltonian $\lambda$. Specifically, as metric we choose the DF 1-norm as defined in Eq.~\ref{eq:df_norm} with $N_{\text{DF}} = 5N$, a choice previously shown to yield well-converged norms~\cite{rocca2024reducing}. As indicated by Eq.~\ref{eq:df_norm} and demonstrated numerically in Appendix~\ref{appendix:norm}, $\lambda_{\text{DF}}$ increases typically quadratically with $N$. We further show in this appendix that Gaussian basis functions with different exponents contribute unevenly to the 1-norm. Traditionally, basis sets were designed to minimize the number of Gaussian functions to reduce classical computational costs. However, with modern hardware and GPU acceleration, this bottleneck has largely been reduced. Moreover, this design choice no longer aligns with the cost metrics of quantum algorithms, where the Hamiltonian's 1-norm remains a dominant cost driver in resource requirements. Motivated by these shifts in computational priorities, our goal is to enrich existing basis sets by introducing additional Gaussian basis functions to describe each \gls{ao}, without increasing the total number of \glspl{ao}, through simultaneous minimization of both the CISD energy and the Hamiltonian's 1-norm.

Specifically, we optimize the basis set independently for the atoms C, N, O, and F. Starting from cc-pVDZ, we add four basis functions to the $d$ orbitals to match the number of contractions found in the ANO double-zeta basis set (from which cc-pVDZ is originally derived, see Section~\ref{sec:atomic_basis_function}). The cost function we optimize is $g(\bm{\theta}) = (1-\gamma) E_{\text{CISD}}(\bm{\theta}) + \gamma \lambda(\bm{\theta})$, where $\bm{\theta}$ are the contraction coefficients (i.e. $d_{\mu}^i$ in Eq.~\ref{eq:contraction}) and exponents (i.e. $\alpha_i$ in Eq.~\ref{eq:contraction}) of the basis functions to optimize in the basis set. The parameter $\gamma$ is varied, and the optimal basis corresponds to the $\gamma$ value yielding the best performance. The energy term $E_{\text{CISD}}(\bm{\theta})$ corresponds to the smallest molecule containing the atom of interest and hydrogen, namely CH\textsubscript{4}, NH\textsubscript{3}, H\textsubscript{2}O, and HF for C, N, O, and F, respectively, taken at equilibrium geometry. 

The optimization landscape contains numerous local minima, making it crucial to select initial parameters carefully. We outline our procedure using CH\textsubscript{4} as an example. The original cc-pVDZ contraction for the $d$ orbitals of carbon is given in Table~\ref{tab:C_basis}.
\begin{table}[tb]
\centering
\begin{tabular}{lcr}
\toprule
Basis & Exponent ($\alpha_i$) & Contraction ($d_{\mu}^i$) \\
\midrule
cc-pVDZ & 0.5500    & 1.0000000000 \\
\hline 
\multirow{5}{*}{ANO DZ} & 4.5420 & 0.0329025262\\
&1.9790 & 1.0671262673\\
&0.8621 & -0.5964927180\\
&0.3756 & -0.8310168604\\
&0.1636 & 0.9849265660 \\
\hline
\multirow{5}{*}{Initialization} &4.5420 & 0.0000000000 \\
&1.9790 & 0.0000000000 \\
&0.8621 & 0.0000000000 \\
&0.5500 & 1.0000000000 \\
&0.1636 & 0.0000000000 \\
\bottomrule
\end{tabular}
\caption{Basis functions for the $d$-orbitals of Carbon}
\label{tab:C_basis}
\end{table}
It corresponds to one contracted Gaussian (i.e., one \gls{ao}), composed of one primitive Gaussian with an exponent of $0.55$. In contrast, the ANO DZ basis (see Table~\ref{tab:C_basis}) corresponds again to a single contracted Gaussian (a single \gls{ao}), but this time made of 5 primitive Gaussian functions. To align with the ANO contraction structure, we introduce four additional contractions to cc-pVDZ. The ANO exponents are used as starting values, with the one closest to the cc-pVDZ exponent replaced by the cc-pVDZ value. For initialization, we retain only the cc-pVDZ contraction coefficient as non-zero, as shown in Table~\ref{tab:C_basis}. In this way, the starting energy is the cc-pVDZ energy. 

The optimization is performed using the \texttt{LinearConstraint} optimizer from \texttt{SciPy}, with constraints enforcing strictly positive exponents. The basis for hydrogen remains fixed at cc-pVDZ. Optimizations are terminated after 100 steps, at which point convergence to a local minimum is observed. We summarize the results for all tested molecules in Table~\ref{tab:cc-pvdz_opt}. Generally, small improvements in the 1-norm can be achieved without significantly affecting the energy. In most cases, reducing the 1-norm results in a modest deterioration of the energy. For the HF molecule, a more substantial norm reduction was possible while maintaining an energy deviation of less than 1 mHa relative to the original basis.
\begin{table}[tb]
    \centering
    \begin{tabular}{lcrcrc}
    \toprule
        Molecule & $\gamma$ & $E^{\text{CISD}}_{\text{cc-pVDZ}}$ & $\lambda_{\text{cc-pVDZ}}$ & $E^{\text{CISD}}_{\text{opt-dz}}$ & $\lambda_{\text{opt-dz}}$\\
        \hline
        CH\textsubscript{4} & $1.0/\lambda_{\text{cc-pVDZ}}$ &  \multirow{2}{*}{-40.37788} & \multirow{2}{*}{543.5} & -40.37692 & 539.1 \\
        CH\textsubscript{4} & $0.5/\lambda_{\text{cc-pVDZ}}$ &  &  & -40.37708 & 540.0 \\
        \hline
        NH\textsubscript{3} & $0.5/\lambda_{\text{cc-pVDZ}}$ & \multirow{2}{*}{-56.39228} & \multirow{2}{*}{433.3} & -56.38635 & 417.1 \\
        NH\textsubscript{3} & $0.1/\lambda_{\text{cc-pVDZ}}$ &  &  & -56.39819 & 435.5\\
        \hline
        H\textsubscript{2}O & $0.5/\lambda_{\text{cc-pVDZ}}$ & \multirow{2}{*}{-76.23206} & \multirow{2}{*}{328.1} & -76.16579 & 180.8\\
        H\textsubscript{2}O & $0.1/\lambda_{\text{cc-pVDZ}}$ &  &  & -76.23964 & 324.4\\
        \hline
        HF & $0.5/\lambda_{\text{cc-pVDZ}}$ & \multirow{2}{*}{-100.22167} & \multirow{2}{*}{235.1} & -100.14708 & 108.3 \\
        HF & $0.1/\lambda_{\text{cc-pVDZ}}$ &  &  & -100.23126 & 225.6\\
        \bottomrule
    \end{tabular}
    \caption{cc-pVDZ basis set optimization results. We add four primitive Gaussians in the contraction of the $d$ orbitals (to match the number of primitives in the ANO basis set). We optimize the exponents and contraction coefficients, $\bm{\theta}$, in order to minimize $g(\bm{\theta}) = (1-\gamma) E_{\text{CISD}}(\bm{\theta}) + \gamma \lambda(\bm{\theta})$. $\lambda_X$ is the \gls{df} norm in basis set $X$. The energies and norms are given in Hartree.}
    \label{tab:cc-pvdz_opt}
\end{table}

Given that cc-pVDZ is a small basis, we repeat the procedure with cc-pVTZ. Here, both $d$ and $f$ orbitals are optimized. Similarly to the double zeta case, we add primitive Gaussians to cc-pVTZ to match the number of primitives per contracted Gaussian in ANO-TZ, resulting in a basis with five primitives for $d$ and four for $f$ orbitals. The cc-pVTZ basis originally includes two contracted Gaussians for the $d$ orbitals, constructed from the same exponents but with different contraction coefficients. We maintain this structure, leading to 15 parameters to optimize for the $d$ orbitals: 5 exponents and $2\times 5$ contraction coefficients. For each atom, we first optimize the $f$ orbital basis, followed by the $d$ orbital. Here again, the optimization is stopped after 100 steps. The $\gamma$ parameter is smaller than in the double zeta case, to compensate for the larger variations in $\lambda$. To accelerate the optimization, we retain the cc-pVDZ basis for the hydrogen atom and later verify that improvements persist when the cc-pVTZ is used for hydrogen. Table~\ref{tab:cc-pvtz_opt} shows our results. Improvements are generally more significant than in the cc-pVDZ case, except for CH\textsubscript{4}, with reductions in 1-norm of up to 10\% while preserving energy accuracy.
\begin{table}[tb]
    \centering
    \begin{tabular}{lcrcrr}
    \toprule
        Molecule & $\gamma$ & $E^{\text{CISD}}_{\text{cc-pVTZ}}$ & $\lambda_{\text{cc-pVTZ}}$ & $E^{\text{CISD}}_{\text{opt-tz}}$ & $\lambda_{\text{opt-tz}}$\\
        \hline
        CH\textsubscript{4} & $0.05/\lambda_{\text{cc-pVTZ}}$ & -40.41596 & 1435.5 & -40.41561 & 1432.9 \\
        NH\textsubscript{3} & $0.05/\lambda_{\text{cc-pVTZ}}$ & -56.44926 & 1342.7 & -56.45128 & 1208.5 \\
        H\textsubscript{2}O & $0.05/\lambda_{\text{cc-pVTZ}}$ & -76.31158 & 1214.6 & -76.31426 & 1055.0\\
        HF & $0.05/\lambda_{\text{cc-pVTZ}}$ & -100.32686 & 1102.3 & -100.32942 & 960.2 \\
        \bottomrule
    \end{tabular}
    \caption{cc-pVTZ basis set optimization results. We add four and three primitive Gaussians in the contraction of the $d$ and $f$ orbitals, respectively (to match the number of primitives in the ANO basis set). We optimize the exponents and contraction coefficients, $\bm{\theta}$, in order to minimize $g(\bm{\theta}) = (1-\gamma) E_{\text{CISD}}(\bm{\theta}) + \gamma \lambda(\bm{\theta})$. $\lambda_X$ is the \gls{df} norm in basis set $X$. The energies and norms are given in Hartree.}
    \label{tab:cc-pvtz_opt}
\end{table}

Finally, we test the transferability of our optimized basis sets using a molecular dataset containing only H, C, N, O, and F atoms~\cite{ramakrishnan2014quantum}. We select the first 50 molecules in the dataset. Since these first 50 molecules only contain H, C, N, and O atoms, we supplement them with the first eight fluorine-containing molecules in the dataset. These are illustrated in Appendix~\ref{app:data_set}, along with their corresponding indices in the dataset.

\begin{figure}[tb]
    \centering
    \begin{subfigure}[b]{0.49\linewidth}
        \centering
        \includegraphics[width=\linewidth]{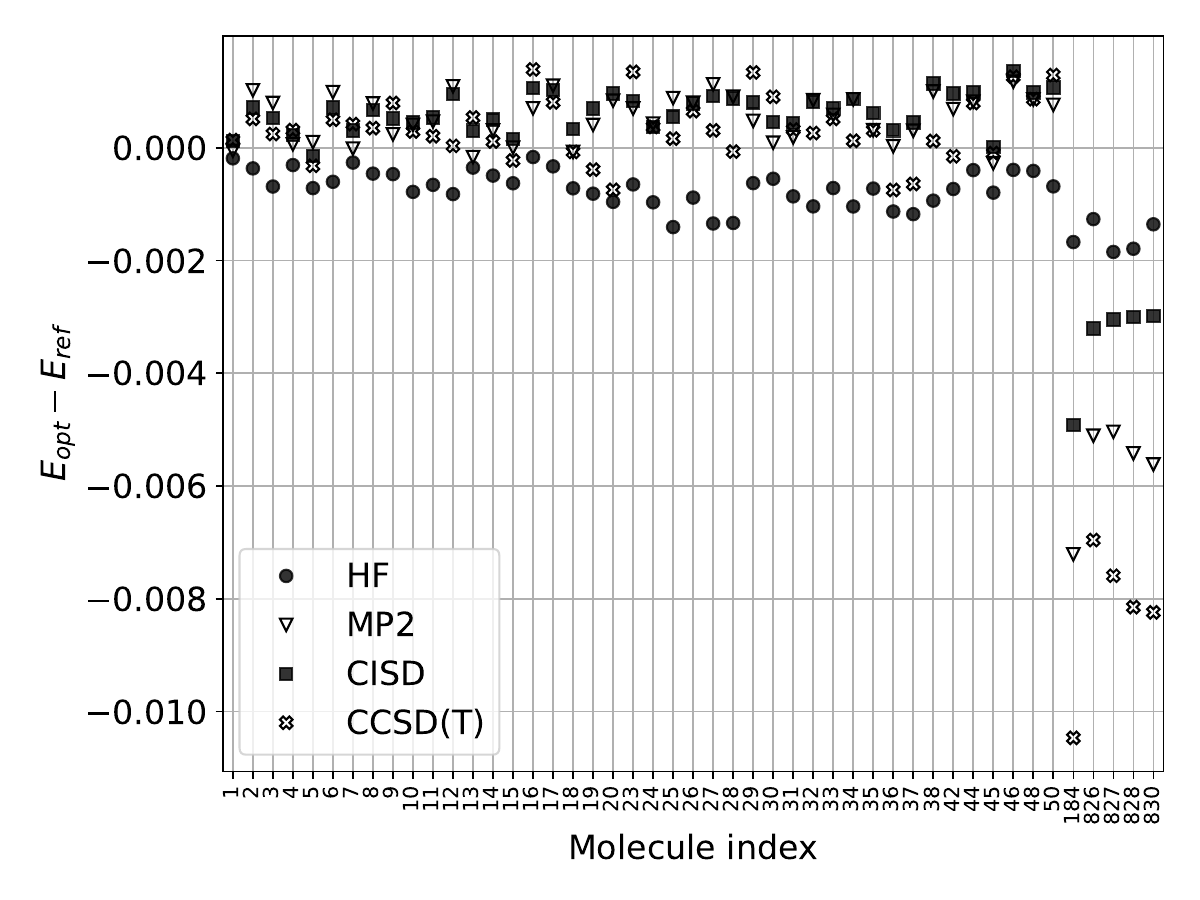}
        \caption{}
        \label{fig:opt_TZ_a}
    \end{subfigure}
    \hfill
    \begin{subfigure}[b]{0.49\linewidth}
        \centering
        \includegraphics[width=\linewidth]{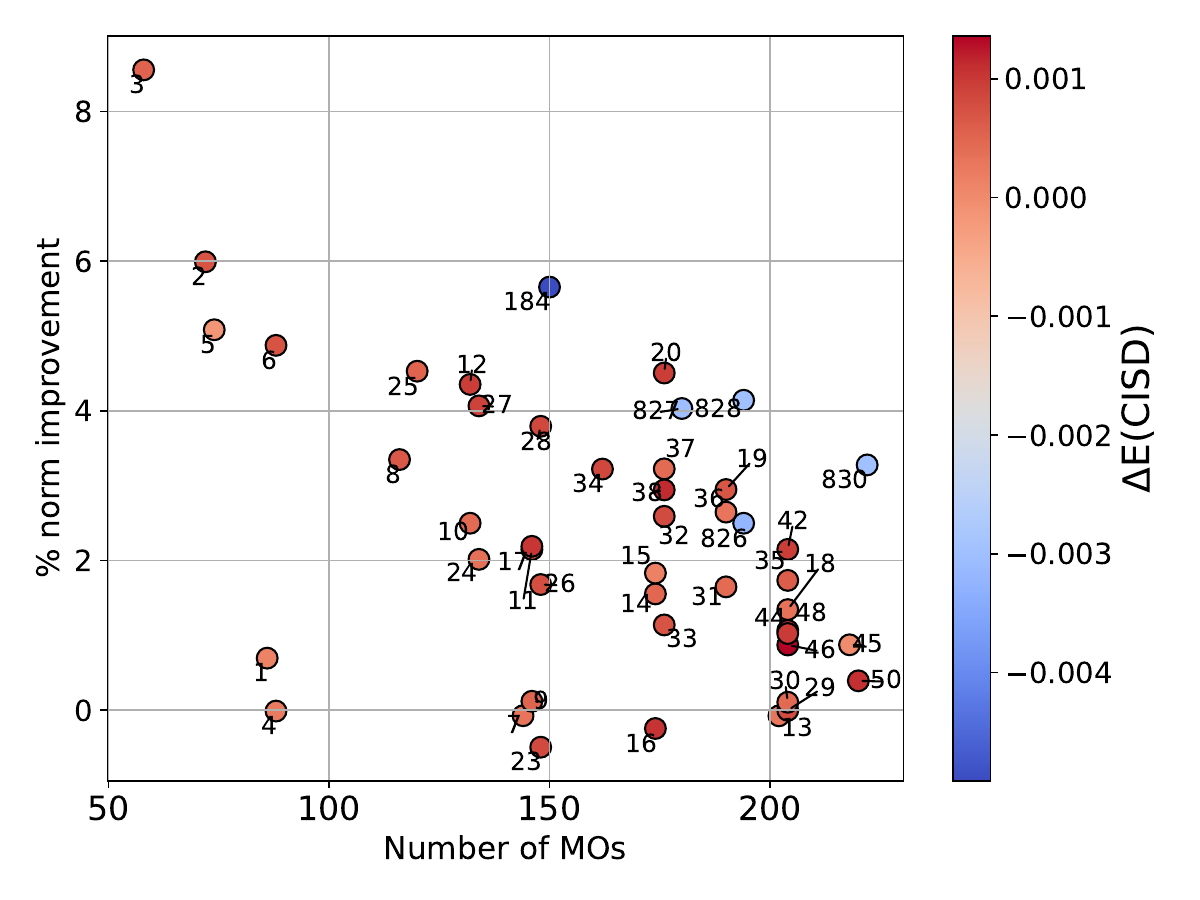}
        \caption{}
        \label{fig:opt_TZ_b}
    \end{subfigure}

    \caption{(a) Differences between the energy calculated in cc-pVTZ and its optimized counterpart at different levels of theory. Filled markers denote variational methods, and empty markers denote non-variational methods. Data points are obtained for molecules in the dataset represented in Fig.~\ref{fig:skeletal}. The corresponding molecule index is given on the x-axis. The optimized basis set is obtained by starting from cc-pVTZ and incorporating additional contractions for the $d$ and $f$ orbitals. The exponents and coefficients for these orbitals are optimized to minimize both the CISD energy and the \gls{df} 1-norm. The full procedure is described in the main text. (b) Percentage norm improvement obtained by using the optimized version of the cc-pVTZ basis set versus the number of orbitals for molecules in the dataset represented in Fig.~\ref{fig:skeletal}. The molecule index is given near its marker. The marker color indicates the difference in CISD energy (Ha) obtained in the optimized and reference basis set, i.e. $\Delta E\text{(CISD}) = E^{\text{opt}}\text{(CISD)} - E^{\text{ref}}\text{(CISD)}$}
    \label{fig:optimized_TZ}
\end{figure}

For each molecule, we compute the \gls{hf}, \gls{mp2}, CISD, and CCSD(T) energies and corresponding DF 1-norm using both the original cc-pVTZ basis and our optimized basis set for all atoms except H (which remains cc-pVTZ). Molecules with more than 230 \glspl{mo} are excluded. The computed energies are displayed in Fig.~\ref{fig:opt_TZ_a}. For molecules containing no fluorine, the correlation energy tends to worsen slightly in the optimized basis. The percent improvement in 1-norm, plotted in Fig.~\ref{fig:opt_TZ_b}, remains below 10\% for all molecules and generally decreases with increasing system size. 

Our findings suggest that while our optimization strategy can reduce the DF 1-norm, its impact is limited, particularly for larger systems. This highlights the challenge of achieving meaningful resource savings without compromising accuracy. 

\subsection{Truncation of molecular orbitals using frozen natural orbitals}
\label{sec:res:fno_strategy}

In this section, we explore an alternative approach for reducing the resource requirements of QPE, exploiting frozen natural orbitals (FNOs). Unlike the optimization of the basis set parameters, FNO truncates the virtual orbital space by leveraging occupation number information derived from correlated wavefunctions.
This is an effective approach for reducing the Hamiltonian 1-norm and block-encoding cost, while preserving the accuracy in estimating the correlation energy. In this section, we quantify the impact of FNO truncation on the DF 1-norm and orbital count, and benchmark its effectiveness across different basis sets and molecular systems.

As a metric, we are using the correlation energy, defined as
\begin{equation}
    E_{\text{corr}}^{\text{ORB,BS}} = E_{\text{CCSD(T)}}^{\text{ORB,BS}} - E_{\text{HF}}^{\text{BS}}\,.
    \label{eq:correlation_energy}
\end{equation}
where BS denotes the basis set used—such as cc-pVDZ (DZ), cc-pVTZ (TZ), or cc-pVQZ (QZ). The CCSD(T) energy is evaluated in a specific molecular orbital basis (ORB), which may correspond to canonical \gls{hf} orbitals (MO) or to \glspl{fno} (NO).

\begin{figure}[tb]
    \centering
    \includegraphics[width=1.0\linewidth]{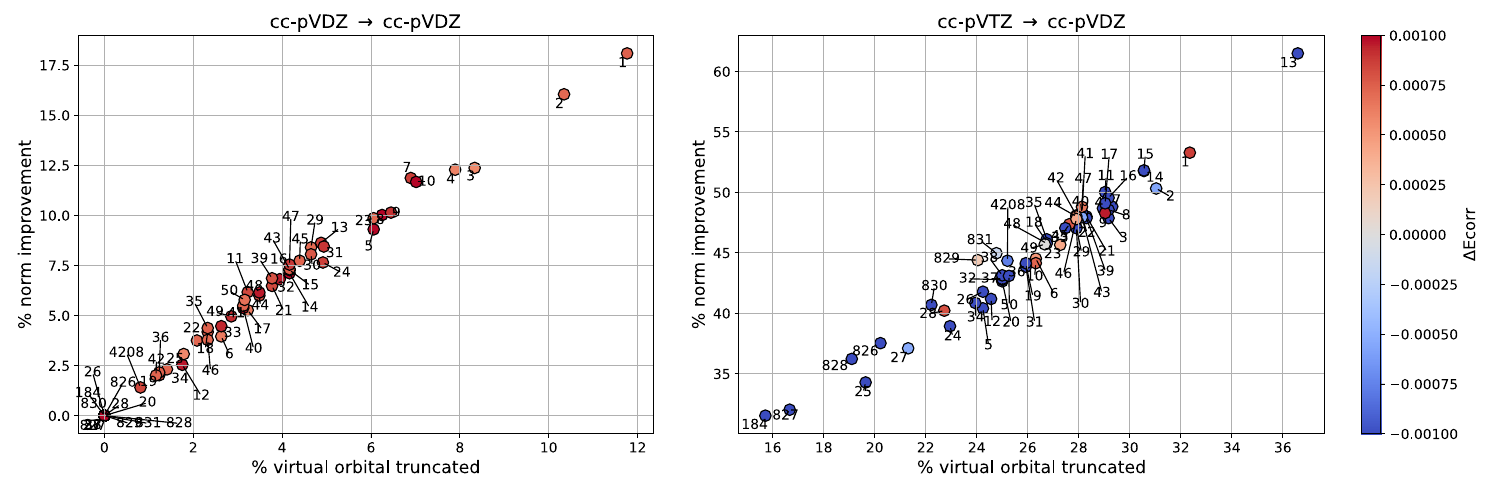}
    \caption{Improvement in the \gls{df} norm obtained by truncating the \gls{mp2} \glspl{no} virtual space. The \gls{mp2} \glspl{no} virtual space is either obtained in the cc-pVDZ (left) or in the cc-pVTZ (right) basis. The percentage improvements in both the norm and the number of orbitals are with respect to the cc-pVDZ basis set. The data is obtained for molecules in the set represented in Fig.~\ref{fig:skeletal}. The molecule index is given near its marker. The marker color indicates the difference in correlation energy (Ha) obtained from the CCSD(T) energy in the remaining NO space and the canonical CCSD(T)/cc-pVDZ (see Eq.~\ref{eq:correlation_energy} and main text). }
    \label{fig:dz_no}
\end{figure}

Following the \gls{fno} strategy, we compute the \gls{mp2} natural orbitals for each small molecule in our dataset (see Appendix~\ref{app:data_set}) using the cc-pVDZ basis. We then truncate the virtual space by removing as many orbitals as possible while ensuring that: $E_{\text{corr}}^{\text{NO,DZ}} - E_{\text{corr}}^{\text{MO,DZ}} < 1 \text{mHa}$. 
Figure~\ref{fig:dz_no} (left) displays the percentage improvement in the norm as a function of the percentage of truncated virtual orbitals. A clear linear relationship emerges: larger truncations yield greater norm reductions, with observed improvements ranging from 0\% to 18\%.
In the context of \gls{qpe}, this strategy yields dual benefits: reducing the number of orbitals decreases the cost of block encoding. In contrast, the lowered 1-norm reduces the number of required walk operator applications. Both effects contribute to more efficient quantum simulations.

We extend this approach further by using a larger basis set, cc-pVTZ, as the starting point for the \gls{fno}. After obtaining the \gls{mp2} \glspl{no} in this larger basis, we truncate them such that the final correlation energy approximates that of the canonical CCSD(T) energy in cc-pVDZ, i.e., $E_{\text{corr}}^{\text{NO,TZ}} - E_{\text{corr}}^{\text{MO,DZ}} < 1 \text{mHa}$. For clarity we also explicitly rewrite $E_{\text{corr}}^{\text{NO,TZ}} = E_{\text{CCSD(T)}}^{\text{NO,TZ}} - E_{\text{HF}}^{\text{TZ}}$.

This yields significantly greater improvements, as shown in Fig.~\ref{fig:dz_no} (right), with norm reductions between 30\% and 60\%, and virtual space reductions between 16\% and 36\%. We look at the occupation number of the last virtual \gls{no} included in the orbital space for each molecule. The mean occupation numbers are $\text{NOON}_{\text{DZ}\rightarrow\text{DZ}} = (1.19 \pm 0.713) \times 10^{-4}$ and $\text{NOON}_{\text{TZ}\rightarrow\text{DZ}} = (1.14 \pm 0.204) \times 10^{-3}$. 

\begin{figure}[tb]
    \centering
    \includegraphics[width=1.0\linewidth]{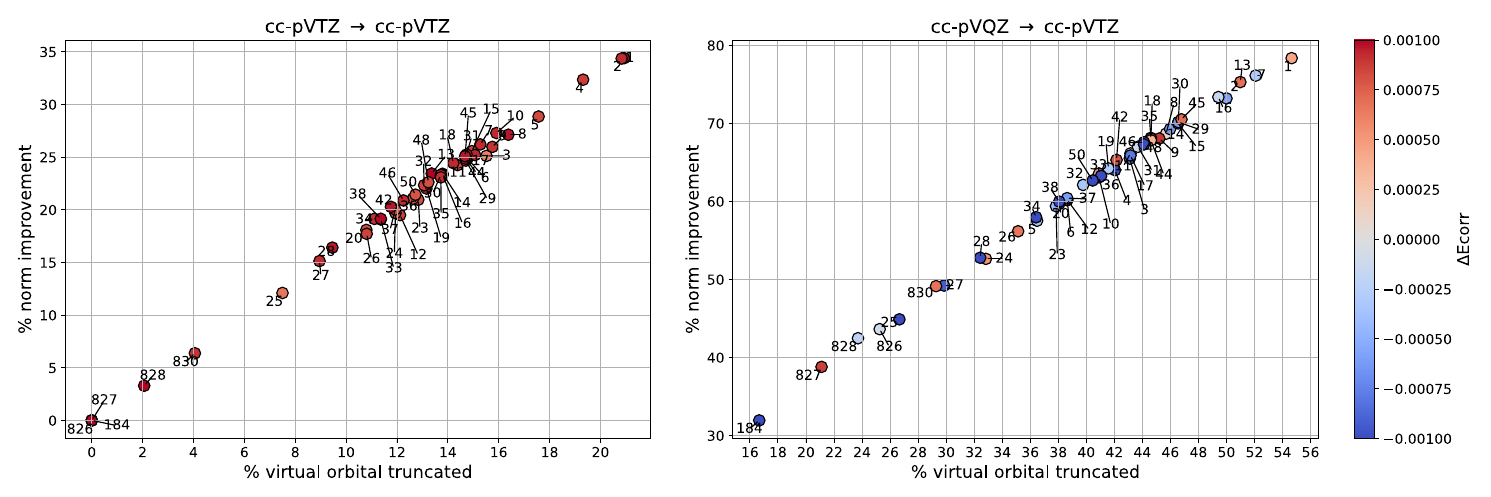}
    \caption{Same as Fig.~\ref{fig:dz_no}, but for the cc-pVTZ basis set. Improvement in the \gls{df} norm obtained by truncating the \gls{mp2} \glspl{no} virtual space. The \gls{mp2} \glspl{no} virtual space is either obtained in the cc-pVTZ (left) or in the cc-pVQZ (right) basis. The percentage improvements in both the norm and the number of orbitals are with respect to the cc-pVTZ basis set. The data is obtained for molecules in the set represented in Fig.~\ref{fig:skeletal}. The molecule index is given near its marker. The marker color indicates the difference in correlation energy (Ha) obtained from the CCSD(T) energy in the remaining NO space and the canonical CCSD(T)/cc-pVTZ (see Eq.~\ref{eq:correlation_energy} and main text). }
    \label{fig:tz_no}
\end{figure}

Finally, we repeat the same procedure but this time calculating the \glspl{fno} in cc-pVTZ and cc-pVQZ, aiming to recover $E_{\text{corr}}^{\text{MO,TZ}}$. Here, the improvements, shown in Fig.~\ref{fig:tz_no}, spread over a broader range and can reach up to almost 80\% in norm improvement and 55\% in the number of orbitals. 

These results suggest that, for a fixed target accuracy, computing low-accuracy correlation in a large basis followed by truncation of the virtual space based on NO occupation provides a more compact and efficient orbital set than doing this procedure directly in a small basis set. The primary computational bottleneck is the \gls{mp2} calculation in the large basis. Still, once this step is completed, the resulting compressed orbital space enables substantial acceleration of the subsequent quantum simulation. 

We also study a case where static correlation plays an important role. We examine the dissociation curve of N$_2$. For reference, we calculate CASSCF+NEVPT2 energies using an active space of 10 electrons in 12 orbitals. These reference dissociation curves are obtained in both the cc-pVDZ and cc-pVTZ basis sets.

We then calculate \gls{mp2} energies and \glspl{no} in both basis sets. This yields two sets of \glspl{fno}: one from the cc-pVDZ basis set obtained with a truncation threshold of $\sigma_{\mathrm{NO}} = 10^{-4}$, and one from the cc-pVTZ basis set obtained with $\sigma_{\mathrm{NO}} = 10^{-3}$. These thresholds were selected based on previous observations of the mean occupation numbers found in Section~\ref{sec:res:basis_optimization}. Using both sets of \glspl{fno}, we perform \gls{sci} calculations, as implemented in Dice\cite{sharma2017semistochastic, holmes2016heat}, with a variational threshold $\epsilon = 10^{-4}$ and no perturbative correction, ensuring variational consistency.

\begin{figure}[tb]
    \centering
    \begin{subfigure}[b]{0.49\linewidth}
        \centering
        \includegraphics[width=\linewidth]{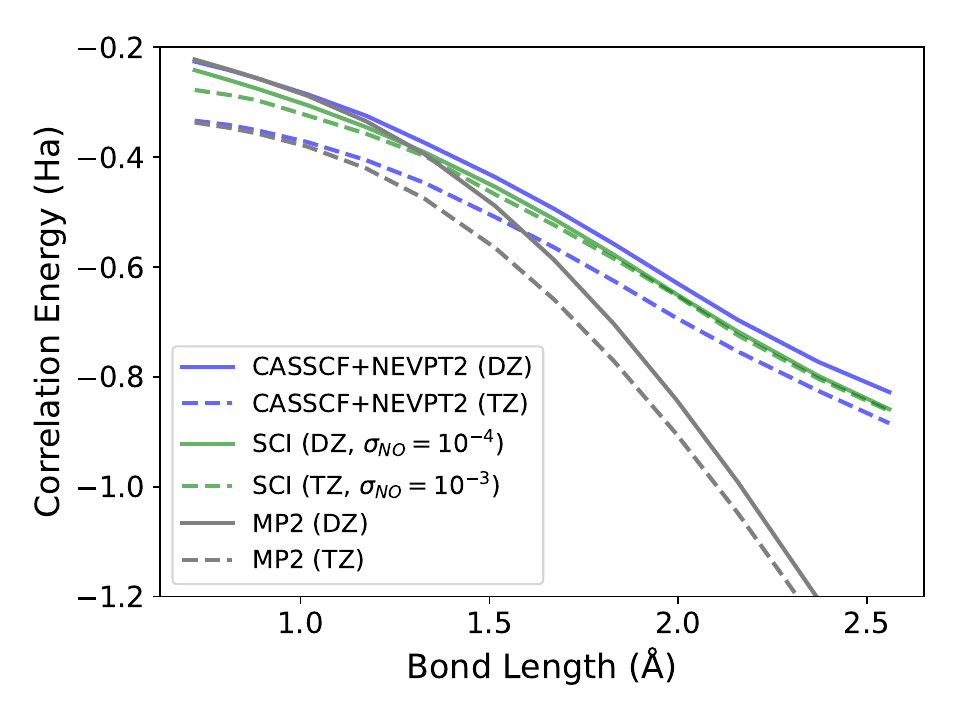}
        \caption{}
        \label{fig:N2_dissoc_a}
    \end{subfigure}
    \hfill
    \begin{subfigure}[b]{0.49\linewidth}
        \centering
        \includegraphics[width=\linewidth]{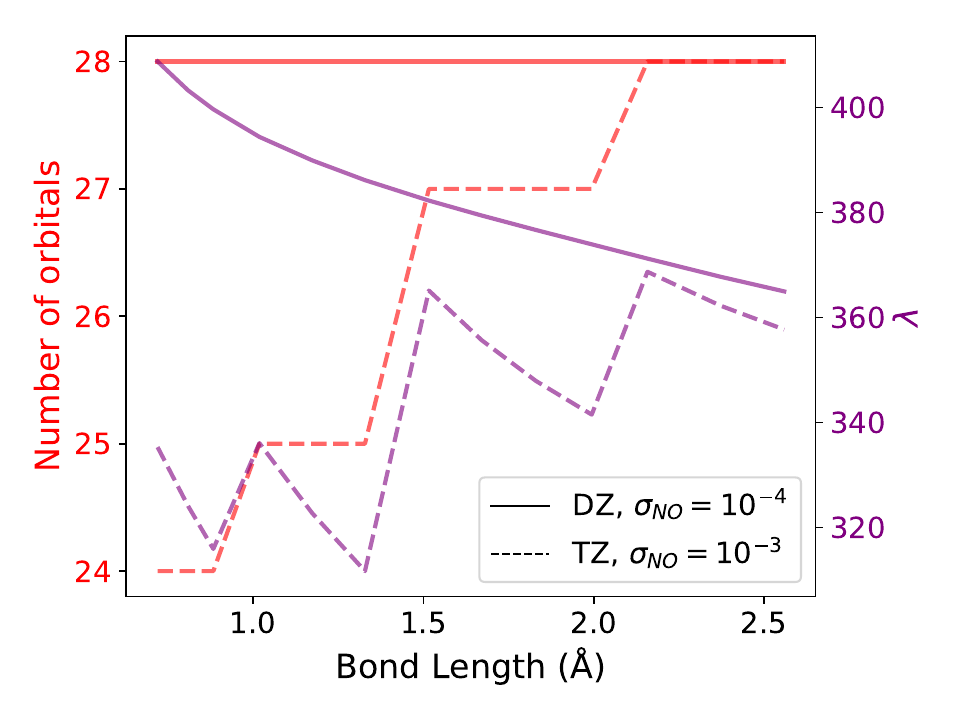}
        \caption{}
        \label{fig:N2_dissoc_b}
    \end{subfigure}

    \caption{N$_2$ dissociation. (a) Correlation energy versus the bond length of the N$_2$ molecule. The blue lines are obtained from a CASSCF in an active space with 10 electrons in 12 orbitals for both the cc-pVDZ and the cc-pVTZ basis sets. The green lines are obtained by running a SCI with a variational threshold $\epsilon_1 = 10^{-4}$ and no perturbative step, in the truncated \gls{no} space obtained from cc-pVDZ and cc-pVTZ at truncation threshold $\sigma_{NO}=10^{-4}$ and $\sigma_{NO}=10^{-3}$, respectively. The grey lines show the \gls{mp2} energies in both basis sets. (b) The corresponding number of \glspl{no} and Hamiltonian (DF) 1-norm, $\lambda$, in both SCI calculations at each bond length. In the cc-pVDZ case, a threshold of $\sigma_{NO}=10^{-4}$ includes 28 \glspl{no} at every bond length, and the associated 1-norm decreases smoothly with increasing bond length. On the other hand, for cc-pVTZ with $\sigma_{NO}=10^{-3}$, the number of included \glspl{no} grows as the bond is stretched, causing the jagged features in the corresponding one-norm curve.  The associated \gls{qpe} cost of the system obtained from cc-pVTZ and $\sigma_{NO}=10^{-3}$ is always better than the other case and always leads to slightly better correlation energies.}
    \label{fig:N2_dissoc}
\end{figure}

The results are presented in Fig.~\ref{fig:N2_dissoc}. First, we observe that the \gls{mp2} correlation energies rapidly become unreliable in both basis sets, significantly overestimating the correlation energy, particularly at stretched geometries. In contrast, the \gls{sci} calculations using \glspl{fno} recover the correct qualitative behavior. The resulting correlation energies are slightly better than those obtained from CASSCF in the cc-pVDZ basis. Moreover, the energies obtained from the cc-pVTZ \glspl{fno} are consistently superior, especially at short bond lengths where they reproduce the shape of the CASSCF curve in cc-pVTZ, something that cc-pVDZ fails to describe adequately.

Importantly, the number of orbitals and the corresponding $\lambda$ values are always lower (or equal) for the system built from cc-pVTZ \glspl{fno}. This suggests that, along the dissociation curve, even when \gls{mp2} fails, achieving cc-pVDZ-level accuracy is more efficient by performing \gls{qpe} using \glspl{fno} derived from the cc-pVTZ basis set, rather than using \glspl{fno} obtained directly from the cc-pVDZ basis. In this case, the \gls{mp2} energy in the large basis set is inaccurate and a subsequent calculation at the \gls{fci} level (or approximation thereof), even at smaller-basis-set quality, is warranted. 

As a final step, we verify the convergence of $\lambda$ with respect to $N_{\text{DF}}$ (see Eq.~\ref{eq:df_2electron_tensor}). For both the (DZ, $\sigma_{\text{NO}} = 10^{-4}$) and (TZ, $\sigma_{\text{NO}} = 10^{-3}$) cases, we find that $\lambda$ is converged to milli-Hartree accuracy for $N_{\text{DF}} = 7N$ across the entire dissociation curve. This result highlights that the block-encoding cost $C_W$ is effectively governed solely by $N$, and further underscores the superior performance of the TZ \gls{fno} basis.

\section{Conclusion}
\label{sec:conclusion}

In this work, we investigated strategies to reduce the computational cost of the \gls{qpe} algorithm by optimizing the basis set. Such optimization becomes particularly important when describing large active spaces or full orbital spaces, where dynamic correlation must be accurately captured and calculations approach the basis set limit.

As a first strategy, we explored the optimization of the Gaussian basis function coefficients to minimize the 1-norm of the Hamiltonian for small organic molecules composed of H, C, N, O, and F atoms. While our approach led to improvements up to $\sim$10\%  reduction in the Hamiltonian's 1-norm, our numerical results showed that our method is system-dependent and offered limited practical gains, particularly when the number of molecular orbitals increased. These results suggest that while the tuning of basis functions can contribute to the reduction of costs, it is not sufficient on its own to scale \gls{qpe} computations towards the basis set limit.

Our second strategy focused on how the choice of basis set affects the number of orbitals required for a given target accuracy, since the cost of \gls{qpe} mostly depends on the number of orbitals. Using a dataset of 58 small organic molecules and the dissociation curve of N$_2$, we demonstrated that for a fixed target accuracy, it is more efficient to compute \glspl{fno} in a large, over-performing basis set and then truncate the orbital space to match the required lower accuracy, rather than using a smaller basis set from the start. This strategy led to reductions in the number of retained orbitals by up to 55\%, and Hamiltonian 1-norm improvements of up to 80\%, resulting in significantly lower \gls{qpe} resource requirements.

Looking ahead, there is potential to obtain \glspl{no} from methods other than \gls{mp2}, such as low bond dimension \gls{dmrg}~\cite{stein2019autocas}, which may yield improved orbitals in regimes where \gls{mp2} breaks down. Moreover, our large basis set \gls{fno} strategy could be integrated into an extrapolation scheme, as the associated errors decrease linearly with the total recovered natural occupation~\cite{landau2010frozen}. This approach could enable the recovery of large basis set accuracy at significantly lower \gls{qpe} cost, requiring only a few repetitions of the \gls{qpe} algorithm. 

Lastly, in this work, we focused exclusively on the effect of the basis set within the \gls{fno} framework on total energy, given our emphasis on \gls{qpe}. Certainly,  the convergence of other molecular properties, such as nuclear forces and dipole moments, needs to be studied in the future. These properties should also exhibit favorable convergence behavior for the method to be broadly applicable.

\begin{acknowledgement}
The authors thank Clemens Utschig-Utschig for his comments on the manuscript and his support during the project. The authors also thank Rob Parrish for useful discussions. P.J.O., J.F.G., and D.R. own stock/options in QC Ware Corp.
\end{acknowledgement}

\providecommand{\latin}[1]{#1}
\makeatletter
\providecommand{\doi}
  {\begingroup\let\do\@makeother\dospecials
  \catcode`\{=1 \catcode`\}=2 \doi@aux}
\providecommand{\doi@aux}[1]{\endgroup\texttt{#1}}
\makeatother
\providecommand*\mcitethebibliography{\thebibliography}
\csname @ifundefined\endcsname{endmcitethebibliography}  {\let\endmcitethebibliography\endthebibliography}{}

\appendix

\section{Norm scaling and Gaussian width dependence}
\label{appendix:norm}

\begin{figure}[h!]
    \centering
    \includegraphics[width=1.0\linewidth]{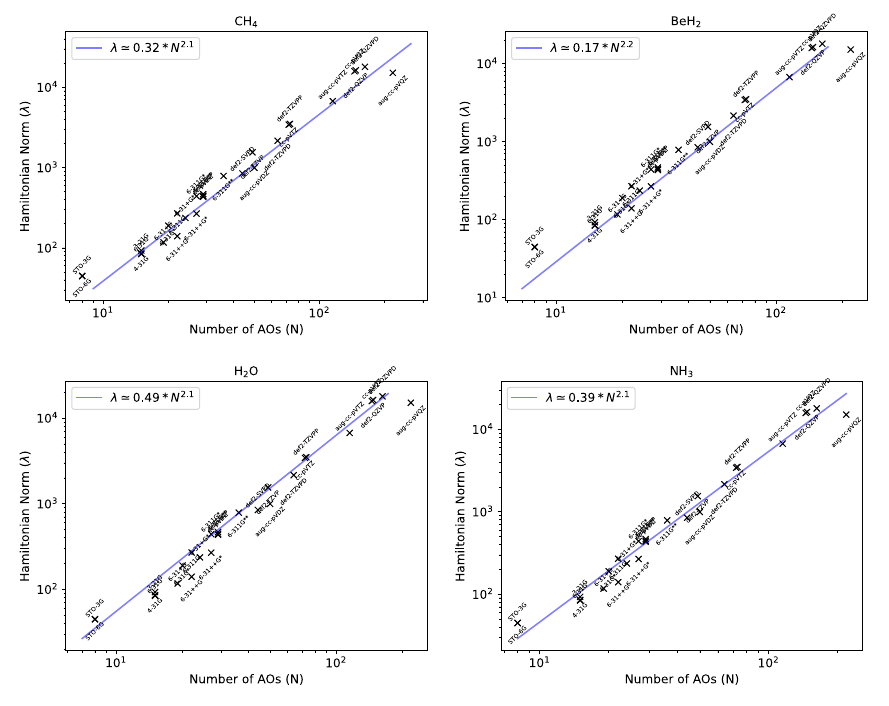}
    \caption{Scaling of the \gls{df} norm ($\lambda$) with the number of \glspl{ao} (N) for CH$_4$, BeH$_2$, H$_2$O and NH$_3$. The data points are obtained for a range of existing basis sets, whose names are listed on the plots. The scaling of $\lambda$ with respect to N is also reported.}
    \label{fig:lambda_vs_nao}
\end{figure}

In this section, we numerically verify in Fig.~\ref{fig:lambda_vs_nao} that the 1-norm (understood throughout as the DF norm of Eq.~\ref{eq:df_norm} with $N_{\text{DF}}=5N$, unless stated otherwise) increases with the number of \glspl{mo}, with almost quadratic scaling. This is demonstrated by computing the norm across various basis sets for four small molecules. Importantly, even for small molecules, using large basis sets results in a significantly increased 1-norm.

\begin{figure}[h!]
    \centering
    \includegraphics[width=1.0\linewidth]{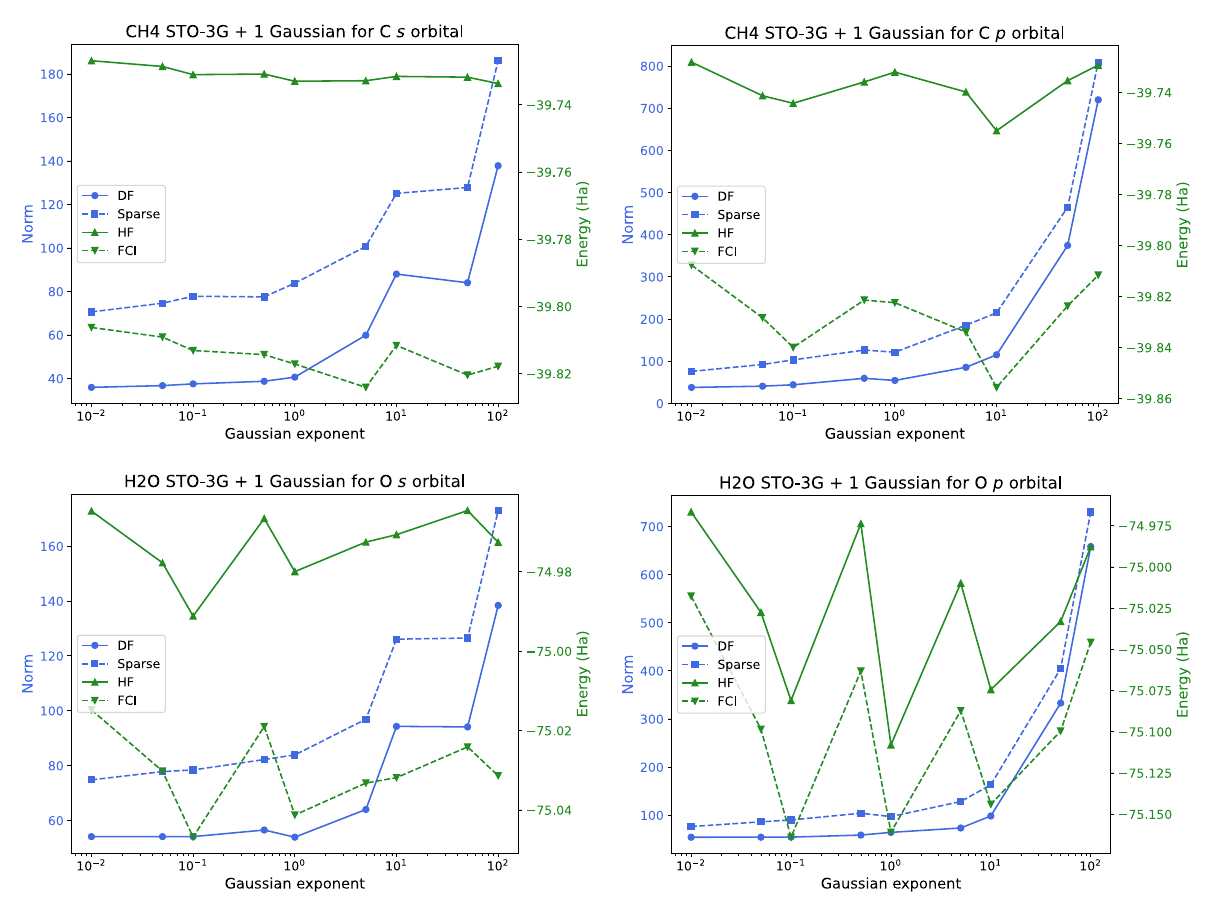}
    \caption{Hamiltonian norm and ground state energy obtained by adding an extra primitive Gaussian to the STO-3G basis set. This is shown for CH$_4$ and H$_2$O as well as for the Gaussian added to the heavy atom's $s$ or $p$ orbitals. The energies are calculated using \gls{hf} and \gls{fci}. Both the sparse (Eq.~\ref{eq:sparse_norm}) and \gls{df} (Eq.~\ref{eq:df_norm}) norms are reported.}
    \label{fig:gaussian_exponent}
\end{figure}

Moreover, Figure~\ref{fig:gaussian_exponent} further illustrates that adding a Gaussian function with a large exponent deteriorates the 1-norm more substantially. Thus, when augmenting basis sets, a trade-off arises between improving the energy and maintaining a manageable 1-norm.

\section{Molecular data set}
\label{app:data_set}

\begin{figure}[h!]
    \centering
    \includegraphics[width=1.0\linewidth]{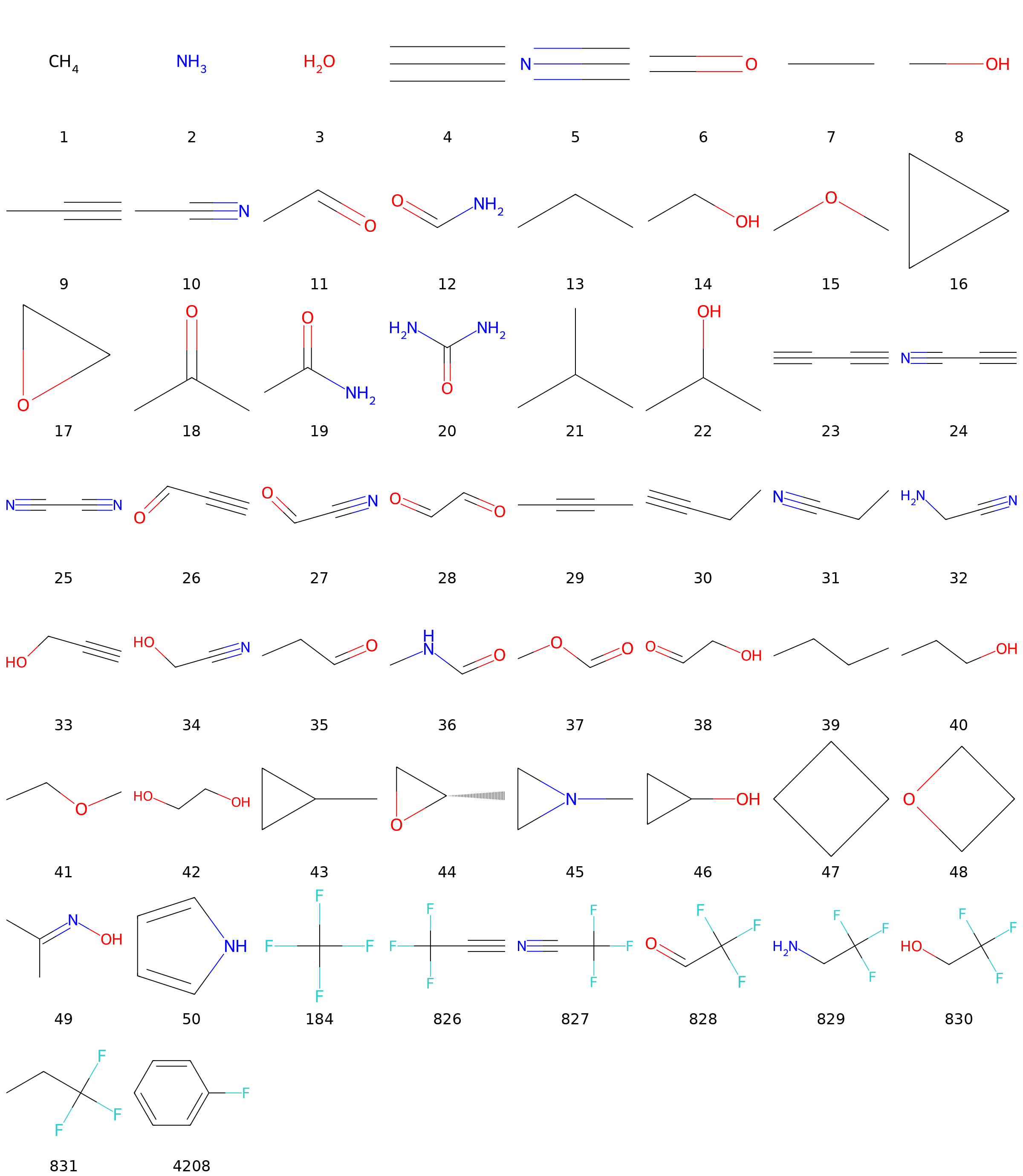}
    \caption{Molecular dataset. The label number corresponds to the molecule index in the original dataset of Ref.~\cite{ramakrishnan2014quantum}.}
    \label{fig:skeletal}
\end{figure}

The molecular dataset used throughout this work is shown in Fig.~\ref{fig:skeletal}. It is constructed from Ref.~\cite{ramakrishnan2014quantum} by taking the first 50 molecules of the original dataset (these 50 molecules contain only H, C, N, and O atoms), and supplementing them with the first eight fluorine-containing molecules in the original dataset.

\end{document}